\documentclass[a4paper,12pt]{article}

\usepackage{amsmath,amssymb,booktabs}
\usepackage{epsf,epsfig}
\usepackage{psfrag}
\usepackage{cite}
\usepackage{mathrsfs}
\usepackage{bbm}
\usepackage[caption=false]{subfig}
\usepackage{dcolumn}
\usepackage[export]{adjustbox}
\allowdisplaybreaks

\def\be{\begin{equation}}
\def\ee{\end{equation}}
\def\beq{\begin{eqnarray}}
\def\eeq{\end{eqnarray}}

\def\DB0{\partial B_0}

\def\Cl2{\mbox{Cl}_2}

\usepackage{color}

\definecolor{Brown}{rgb}{0.5,0.25,0}

\begin{document}



\begin{center}
\Large{\bf\boldmath
Discovery potentials of double-charm tetraquarks
\unboldmath}

\normalsize
\vskip 0.8cm

{ Qin~Qin\footnote{qqin@hust.edu.cn, corresponding author},}
{ Yin-Fa Shen}
\vskip 0.2cm
{\it School of physics, Huazhong University of Science and Technology, Wuhan 430074, China}\\
\vskip 0.5cm
{ Fu-Sheng Yu\footnote{yufsh@lzu.edu.cn, corresponding author}}\\

\vskip 0.2cm

{\it School of Nuclear Science and Technology,  and Frontiers Science Center for Rare Isotopes, Lanzhou University, Lanzhou 730000, China}\\
{\it Lanzhou Center for Theoretical Physics, and Key Laboratory of Theoretical Physics of Gansu Province, Lanzhou University, Lanzhou 730000, China}\\
{\it Center for High Energy Physics, Peking University, Beijing 100871, China}
\vskip 0.8cm

\end{center}


\begin{abstract}
In this study, we investigate the discovery potential of 
double-charm tetraquarks $T^{\{cc\}}_{[\bar{q}\bar{q}']}$. We find that their production cross sections 
at the LHCb with $\sqrt{s} = 13$ TeV reach $\mathcal{O}(10^4)$ pb, which indicates that the LHCb 
has collected $\mathcal{O}(10^8)$ such particles. Through the decay channels of $T^{\{cc\}}_{[\bar{u}\bar{d}]}\to D^{+}K^{-}\pi^{+}$ 
or $D^0D^+\gamma$ (if stable)
or $T^{\{cc\}}_{[\bar{u}\bar{d}]}\to D^0D^{*+}\to D^0D^0\pi^+$ (if unstable), it is highly expected that they get discovered at the LHCb in the near future. 
We also discuss the productions and decays of the double-charm tetraquarks at future Tera-$Z$ factories.
\end{abstract}


\section{Introduction}
\label{sec:intro}

Doubly-heavy tetraquarks including $T^{\{cc\}}_{[\bar{q}\bar{q}']}$, $T^{\{bc\}}_{[\bar{q}\bar{q}']}$, and $T^{\{bb\}}_{[\bar{q}\bar{q}']}$ 
have drawn considerable attention from theorists after the discovery of the first double-charm hadron $\Xi_{cc}^{++}$~\cite{Aaij:2017ueg}. 
The doubly-heavy tetraquarks can shed light on strong dynamics and  greatly help us identify the nature of the exotic $XYZ$ states
or structures, e.g., cusps or true resonances. 
In addition, several studies of both heavy quark symmetry and 
lattice~\cite{Karliner:2017qjm, Eichten:2017ffp, Francis:2016hui, Bicudo:2017szl, Junnarkar:2017sey,Mehen:2017nrh, Czarnecki:2017vco} 
suggest that some of them have masses below the threshold and thus can only decay weakly. If such weakly-decaying 
tetraquarks are found, they will provide evidence of compact diquarks as building blocks of hadronic matter. Among these 
doubly-heavy tetraquarks, $T^{\{cc\}}_{[\bar{q}\bar{q}']}$ is the most promising one to be observed in experiments in the 
near future, because both its production rate and detection efficiency are the highest at the LHCb. In this work, we analyze 
the potential of finding $T^{\{cc\}}_{[\bar{q}\bar{q}']}$ at the LHCb and also future 
$Z$-factories~\cite{CEPCStudyGroup:2018ghi,Gomez-Ceballos:2013zzn}, evaluating their production 
cross sections at the two kinds of facilities and proposing possible golden channels according to their decay properties. 

Decay properties of the doubly-heavy tetraquarks $T^{\{cc\}}_{[\bar{q}\bar{q}']}$ rely on whether their masses are above or 
below the thresholds $D^0D^{*+}$ and $D^*D_s^+$. If they are stable particles below the threshold, they will mainly decay 
radiatively (or even weakly according to \cite{Feng:2013kea}) and thus have a relatively long lifetime, which will 
considerably suppress the background and make the experimental search easier. In contrast, if they are unstable 
particles above the thresholds\footnote{If the particles are slightly below the thresholds, they can also decay strongly and 
thus unstable. For convenience, we also classify this special case as "above threshold".}, 
they will decay strongly and still have a good chance to be found at the LHCb 
using a method similar to that in \cite{Aaij:2019evc}. Their masses have been calculated in many different procedures, such as QCD sum 
rule~\cite{Du:2012wp,Navarra:2007yw,Wang:2017dtg,Agaev:2018vag,Agaev:2019qqn}, lattice~\cite{Francis:2018jyb}, 
Bethe-Salpeter equation approach~\cite{Feng:2013kea}, heavy quark symmetry~\cite{Eichten:2017ffp,Braaten:2020nwp} and several 
QCD inspired models~\cite{Carlson:1987hh,Lee:2009rt,Luo:2017eub,2008.00737,1811.06462,2006.08087,0706.3853,hep-ph/9609348,1911.00215,2004.02106,Karliner:2017qjm,Park:2018wjk,Karliner:2020vsi,Zhu:2019iwm}. The results are listed in Table \ref{tb:masses} in form of 
the mass differences of $T^{\{cc\}}_{\bar{n}\bar{n}'}$ and $T^{\{cc\}}_{\bar{n}\bar{s}}$ ($n^{(\prime)}=u$ or $d$) from their thresholds 
$D^0D^{*+}$ and $D^*D_s^+$, respectively. These results are far from conclusive because some of the calculations suggest they lie above the 
thresholds while the others are against it. As a result, the $T^{\{cc\}}_{[\bar{q}\bar{q}']}$ decay channels are analyzed 
case by case. 

\begin{table}
  \label{tb:masses}
  \centering
  \begin{tabular}[t]{lccccccccccc}\hline\hline
{Reference}  & \cite{Lee:2009rt} & \cite{Luo:2017eub}   & \cite{2008.00737} & \cite{1811.06462} & \cite{2006.08087} & \cite{0706.3853}  & \cite{Du:2012wp} & \cite{Francis:2018jyb} & \cite{Eichten:2017ffp} & \cite{Braaten:2020nwp} & \cite{Wang:2017dtg}\\ \hline
$T^{\{cc\}}_{\bar{n}\bar{n}'}$  & -79 & -96  & +53 & -150 & +166 & +60 & -  & AT  & +102  & +88 & +25 \\ 
$T^{\{cc\}}_{\bar{n}\bar{s}}$  & -9 & -56  & +128 & +94   & +255 & +166 & +143  & AT  & +179  & +181 & -15 \\   \hline\hline
{Reference} & \cite{Feng:2013kea}  & \cite{hep-ph/9609348} & \cite{1911.00215} & \cite{2004.02106} & \cite{Karliner:2017qjm} & \cite{Park:2018wjk} & \cite{Navarra:2007yw} & \cite{Karliner:2020vsi} & \cite{Carlson:1987hh} & \cite{Zhu:2019iwm} \\ \hline
$T^{\{cc\}}_{\bar{n}\bar{n}'}$ & -215  & BT & -149  & -182 & +7 & +98 & +91 &+125  & AT & AT  \\ \hline\hline
\end{tabular}
\caption{Theoretical predictions on the mass differences of $T^{\{cc\}}_{\bar{n}\bar{n}'}$ and $T^{\{cc\}}_{\bar{n}\bar{s}}$ 
from their thresholds $D^0D^{*+}$ and $D^*D_s^+$, respectively, in units of MeV. The quarks $n,n'$ = $u$ or $d$. For 
some works without explicit numerical results, we use "AT" and "BT" denoting "above threshold" and "below threshold",
respectively.}
\end{table}

The rest of the paper is organized as follows. In section \ref{sec:production}, we study the 
production of $T^{\{cc\}}_{[\bar{q}\bar{q}']}$ at both the LHCb and future 
$Z$-factories. According to our estimation of the production cross sections, the LHCb has produced $\mathcal{O}(10^8)$
double-charm tetraquarks to date, and this number will increase by almost one order of magnitude by the end of 
Run 4~\cite{Bediaga:2012py,Bediaga:2012uyd}. It is also shown that a Tera-$Z$ factory will produce $\mathcal{O}(10^6)$ 
double-charm tetraquarks in a very clean environment. 
In section \ref{sec:decay}, possible decay channels of $T^{\{cc\}}_{[\bar{q}\bar{q}']}$
are discussed, in three cases with the double-charm tetraquarks either above or below the $DD^*$ and $DD\gamma$ thresholds. 
For the LHCb, we propose 
some decay channels to search for the double-charm tetraquarks, because they have relatively large branching ratios, and more 
importantly, all of their final-state particles have high detection efficiencies, 
e.g., $T^{\{cc\}}_{[\bar{u}\bar{d}]}\to D^0D^{*+}\to D^0D^0\pi^+$ and $T^{\{cc\}}_{[\bar{u}\bar{d}]}\to D^+K^-\pi^+$. In contrast, future 
$Z$-factories have an advantage over the LHCb in searching for other decay channels containing a photon, e.g., 
$T^{\{cc\}}_{[\bar{u}\bar{d}]}\to D^0D^+\gamma$. We conclude our study in section \ref{sec:conclusion}.

\section{Production rate}
\label{sec:production}

Following the approach proposed in~\cite{Ali:2018ifm,Ali:2019tli,Ali:2018xfq}, we estimate the 
production rate of double-charm hadrons $H_{cc}$ at the LHCb and future $Z$-factories. The basic idea 
is as follows. First, the underlying processes for their production at the quark level are $pp\to cc\bar{c}\bar{c} +X$
and $e^+e^-\to Z\to cc\bar{c}\bar{c}$ at these two kinds of facilities, respectively. The two charm quarks must stay close enough to form 
a $cc$ diquark jet, which further fragments into different kinds of double-charm hadrons. The invariant 
mass of the two heavy quarks $m_{QQ'}$ is used to parameterize their collinear level, and if 
$m_{QQ'}$ is smaller than some cut-off value $M_{QQ'}(\Delta M)\equiv m_Q+m_{Q'}+\Delta M$, we regard the two 
heavy quarks as a $QQ'$ diquark that will eventually produce a doubly-heavy hadron. In contrast,
if $m_{QQ'}$ exceeds $M_{QQ'}(\Delta M)$, the two heavy quarks fragment separately. 
The parameter $\Delta M$ can be determined by matching the partonic $b\bar{c}$ production 
simulation to the experimental measurements of the $B_c$ meson production cross sections \cite{1411.2943}, 
and it has been found by \cite{Ali:2018ifm,Ali:2018xfq} that 
\begin{align}\label{eq:deltam}
\Delta M = \left\{ \begin{array}{l} 
\left( 2.0 ^{+0.5}_{-0.4}\right) \text{GeV, for LHCb,}\\
\left( 2.7 ^{+1.3}_{-0.5}\right) \text{GeV, for Z factories.} \\ \end{array} \right.
\end{align}
Next, we will discuss the production of the doubly-heavy hadrons at the LHCb and future $Z$-factories separately. 

\textit{LHCb:} For the analysis of $H_{cc}$ production at the LHC with $\sqrt{s} = 13$ TeV, we generate $10^5$ 
$pp\to cc\bar{c}\bar{c} +X$ events via the Monte Carlo generator MadGraph5$\_$aMC@NLO~\cite{Alwall:2014hca}
at the next-to-leading-order level showered by Pythia8~\cite{Sjostrand:2006za,Sjostrand:2014zea}. 
It is found that the total cross section $\sigma(pp\to cc\bar{c}\bar{c} +X) \approx 4.9\times 10^7$ pb, 
and after applying the invariant mass cut in \eqref{eq:deltam}, we obtain 
\begin{align}\label{eq:prolhcb}
\sigma(p+p\to H_{cc} + X) = (3.1^{+1.7}_{-0.7}) \times 10^5\ \text{pb} \; ,
\end{align}
in the transverse momentum range $4<p_\text{T}<15$ GeV and the
pseudorapidity range $2< \eta <4.5$. 

\begin{figure}[tb] 
\begin{center}
\includegraphics[width=0.6\textwidth]{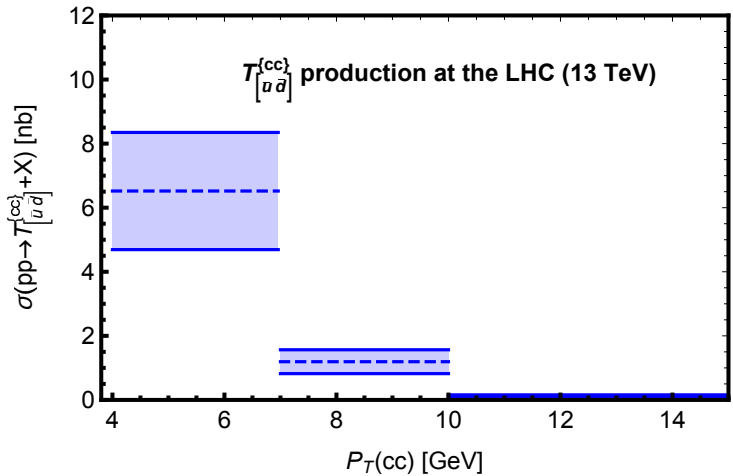} 
\end{center}
\caption{Projected $p_\text{T}$-dependence of the tetraquark production cross section
in $pp \to T^{\{cc\}}_{[\bar u\bar d]} +X$ at the LHC for $\sqrt{s}=13$ TeV with $2<\eta<4.5$.
} \label{fig:ptdist} 
\end{figure}

The double-charm hadrons~$H_{cc}$ include both the double-charm baryons $\Xi_{cc}^{++} (ccu)$, 
$\Xi_{cc}^+ (ccd)$, and $\Omega_{cc}^- (ccs)$ and the double-charm tetraquarks $T^{\{cc\}}_{[\bar q \bar q^\prime]}$
and their excited states. 
Therefore, we still need to know the relative fractions of $H_{cc}$ to these hadrons. It is difficult to calculate these non-perturbative 
fragmentation ratios, while we can approximately borrow the corresponding ratios of a $b$-quark jet under 
the heavy quark - heavy diquark symmetry. Such ratios have been measured by the LHCb~\cite{Aaij:2014jyk,LHCb:2021qbv} as
\begin{equation}\begin{split}\label{eq:frag}
\left[\frac{f_{\Lambda_b^0}}{f_{d}}\right](p_\text{T}) =&\ (0.151 \pm 0.030)+ \exp\left[(-0.57\pm0.11) - (0.095\pm0.016) p_\text{T}\text{(GeV)}\right]\; ,  \\
\left[\frac{f_s}{f_d}\right](p_\text{T}) =&\ (0.263\pm0.008) + ((-17.6\pm2.1)\times 10^{-4})\cdot p_\text{T}\text{(GeV)}\; , 
\end{split}\end{equation}
which are $p_\text{T}$ dependent. 
In addition, we assume that for states containing the same 
valence quarks, the ground state fraction is approximately $r_g = 0.48\pm 0.08$ \cite{Niiyama:2017wpp}. Then, 
convoluting the $p_\text{T}$-distribution of the $cc$-diquark jets fetched from the simulated events with the 
$p_\text{T}$-dependent fragmentation functions \eqref{eq:frag},
we estimate the direct production cross sections of the double-charm tetraquarks to be 
\begin{align}
\sigma(pp\to T^{\{cc\}}_{[\bar u\bar d]} +X) =&\ (24^{+14}_{-7}) \ \text{nb}\; , \nonumber \\
\sigma(pp\to T^{\{cc\}}_{[\bar u\bar s]} +X) =\sigma(pp\to T^{\{cc\}}_{[\bar d\bar s]} +X) =&\ (6.0^{+3.5}_{-1.7}) \ \text{nb}\; .
\end{align}
The $p_\text{T}$ distribution of the $pp\to T^{\{cc\}}_{[\bar u\bar d]} +X$ events from the simulation is displayed in Figure \ref{fig:ptdist}.
In contrast, because almost all the excited states of 
$\Xi_{cc}$ and $\Omega_{cc}$ decay into their ground states, we do not need to multiply the 
$r_g$ ratio for their production cross sections, which read 
\begin{align}\label{eq:xicclhcb}
\sigma(pp\to \Xi_{cc}^{++} +X) = \sigma(pp\to \Xi_{cc}^{+} +X) =&\ (103^{+56}_{-22})\ \text{nb}\; , \nonumber \\
\sigma(pp\to \Omega_{cc}^+ +X) =&\ (26^{+14}_{-6}) \ \text{nb}\; .
\end{align}

The $\Xi_{cc}$ production cross section has been calculated in the framework of non-relativistic QCD as 
$\sigma(\Xi_{cc}^{++}) = \sigma(\Xi_{cc}^+) \approx 62$ nb~\cite{Chang:2006eu}, with the experimental cuts 
$p_\text{T}>$ 4 GeV and $|\eta|\leq1.5$. To compare with this result, we reset the same cuts and find 
$\sigma(\Xi_{cc}^{++}) = \sigma(\Xi_{cc}^+) \approx 100$ nb. Taking into account the large uncertainties, they agree with each other. 
In addition, the $\Xi_{cc}^{++}$
production has been studied by the LHCb, with a relative production rate given as~\cite{Aaij:2019zxa}
\begin{align}\label{eq:promeasure}
{\sigma(\Xi_{cc}^{++})\times \mathcal{B}(\Xi_{cc}^{++} \to \Lambda_c^+K^-\pi^+\pi^+)\over \sigma(\Lambda_c^+)} 
= (2.22 \pm 0.27 \pm 0.29) \times 10^{-4} \; ,
\end{align}
choosing the cuts $4<p_\text{T}<15$ GeV and $2<\eta<4.5$. 
From an LHCb measurement with $\sqrt{s} = 7$ TeV~\cite{Aaij:2013mga}, it can be extracted that 
$\sigma(\Lambda_c^+ + \Lambda_c^-) \approx 38 \mu b$ with $4<p_\text{T}<8\ \text{GeV}, 2.0<\eta<4.5$ (the $p_\text{T}>8$ GeV 
contributions are negligible). By comparing the $D$ meson production at 13 TeV~\cite{Aaij:2015bpa} and 7 TeV~\cite{Aaij:2013mga}, 
one expects that the $\Lambda_c$ production rate with $\sqrt{s} = 13$ TeV should exhibit an enhancement of approximately 50\%, 
{\it i.e.}, $\sigma(\Lambda_c^+ + \Lambda_c^-) \approx 57 \mu b$. Assuming 
$\mathcal{B}(\Xi_{cc}^{++} \to \Lambda_c^+K^-\pi^+\pi^+)\in[5,20]\%$ as suggested in \cite{Yu:2017zst}, 
one can solve from \eqref{eq:promeasure} that $\sigma(\Xi_{cc}^{++})\in [30,130]$ nb, which is also consistent 
with our result given by \eqref{eq:xicclhcb}.

The LHCb collaboration has collected approximately 9 fb$^{-1}$ of data up to now, which
indicates ${\cal O}(10^8)$ $T^{\{cc\}}_{[\bar q\bar q']}$ particles. If reconstructed by proper 
decay channels, as we analyze in the next section, they have a good chance to be discovered in the near future.

\textit{$Z$-factories:} For future $Z$-factories 
which are designed to produce $10^{12}$ or more $Z$ bosons, we also generate $10^5$ 
$e^+e^-\to Z \to cc\bar{c}\bar{c}$ events via MadGraph5~\cite{Alwall:2014hca} and Pythia8~\cite{Sjostrand:2006za,Sjostrand:2014zea}. 
We find that the total cross section $\sigma(e^+e^-\to cc\bar{c}\bar{c} ) \approx 120$ pb, and after 
the diquark jet condition $m_{cc}<M_{cc}(\Delta M)$ is applied, the decay branching ratio of the $Z$ 
boson into double-charm hadrons is found to be 
\begin{align}
\mathcal{B}(Z\to H_{cc} +X ) = (10.1 ^{+7.9}_{-2.7})\times 10^{-5} \; .
\end{align}
Analogous to the analysis of the LHCb, we assume that the fragmentation fraction 
$f ((cc)_\text{jet} \to T^{\{cc\}}_{[\bar u \bar d]} + X)/f ((cc)_\text{jet} \to \Xi_{cc} + X)$ 
is close to two times $f_{\Lambda_b}/(f_{B_u} + f_{B_d}) = 0.11 \pm 0.02$~\cite{Tanabashi:2018oca}, $f_s/f_d\approx 0.25$,
and the ground-state fraction is still $r_g = 0.48\pm 0.08$ \cite{Niiyama:2017wpp}. With these approximations, 
the branching ratios of the $Z$ boson decays into double-charm tetraquarks are estimated to be 
\begin{align}\label{eq:TccZfactory} 
\mathcal{B} (Z \to T^{\{cc\}}_{[\bar u \bar d]} + X) =&\ (4.1^{+3.4}_{-1.5}) \times 10^{-6}\; , \nonumber \\
\mathcal{B} (Z\to T^{\{cc\}}_{[\bar u\bar s]} +X) =\mathcal{B} (Z\to T^{\{cc\}}_{[\bar d\bar s]} +X) =&\ (1.0^{+0.8}_{-0.4}) \times 10^{-6}\; .
\end{align}
We also give the $Z$ decay branching ratios to the ground-state double-charm baryons, 
including those secondarily decaying from their excited states, 
\begin{align}\label{eq:XiccZfactory} 
\mathcal{B} (Z\to \Xi_{cc}^{++} +X) = \mathcal{B} (Z\to \Xi_{cc}^{+} +X) =&\ (3.9^{+3.1}_{-1.1}) \times 10^{-5}\; , \nonumber \\
\mathcal{B} (Z\to \Omega_{cc}^+ +X) =&\ (1.0^{+0.8}_{-0.3}) \times 10^{-5}\; .
\end{align}
It can be seen that our estimation for the $\Xi_{cc}$ production rate at $Z$-factories is consistent with the NRQCD 
calculation $\mathcal{B} (Z\to \Xi_{cc} +X)\approx (1.7^{+1.3}_{-0.6}) \times 10^{-5}$~\cite{Jiang:2012jt}.
Therefore, it is expected that future Tera-$Z$ factories will produce $\mathcal{O}(10^6)$ $T^{\{cc\}}_{[\bar{q}\bar{q}']}$ particles. 
Owing to the clean background of electron-positron collisions, future $Z$-factories have the advantage of 
measuring some of the decay channels of $T^{\{cc\}}_{[\bar{q}\bar{q}']}$ containing 
neutral particles such as photons in their final states, while it is difficult for hadron colliders. More details are given in the next 
section.

\section{Decay channels}
\label{sec:decay}
In addition to the production rates, the decaying processes are another important issue in the experimental 
searches for the double-charm tetraquarks. An example of the importance of decaying processes is the discovery 
of the first doubly charmed baryon. With the prediction of the most favorable decay channels of 
$\Xi_{cc}^{++}\to \Lambda_c^+K^-\pi^+\pi^+$ and $\Xi_c^+\pi^+$ \cite{Yu:2017zst}, the LHCb collaboration 
observed $\Xi_{cc}^{++}$ via the above two modes~\cite{Aaij:2017ueg,LHCb:2018pcs}. 

The decaying properties depend on the masses of $T^{\{cc\}}_{[\bar{q}\bar{q}']}$. 
Above the $DD^*$ thresholds, they will decay strongly. 
In contrast, between the $DD^*$ and $DD\gamma$ thresholds, they will decay radiatively; 
below the $DD\gamma$ thresholds, they will decay weakly. 
Since the theoretical predictions on the masses are very different from each other, as shown in Table~\ref{tb:masses}, 
we will discuss the decay processes separately according to the above cases. 

In the case that the masses of double-charm tetraquarks are above the thresholds of $DD^*$, 
they are unstable and will decay strongly. The decaying processes include 
$T^{\{cc\}}_{[\bar{u}\bar{d}]}\to D^0D^{*+}~\text{or}~D^+D^{*0}$, $T^{\{cc\}}_{[\bar{u}\bar{s}]}\to D^0D_s^{*+}~\text{or}~D_s^+D^{*0}$, and $T^{\{cc\}}_{[\bar{d}\bar{s}]}\to D^+D_s^{*+}~\text{or}~D_s^+D^{*+}$.
The reconstructions of $D^{*0}$ and $D_s^{*+}$ require the reconstructions of neutral particles $\pi^0$ or $\gamma$, 
which will reduce the detection efficiencies significantly. 
Then, the favorable decay processes of the LHCb are 
\begin{align}
&T^{\{cc\}}_{[\bar{u}\bar{d}]}\to D^0D^{*+} \to D^0 D^0 \pi^+,~~~\text{and}~~~T^{\{cc\}}_{[\bar{d}\bar{s}]}\to D_s^+D^{*+}\to D_s^+D^0\pi^+.
\end{align}

If their masses are above the $DD\gamma$ thresholds but below the $DD^{*}$ thresholds, 
they will decay radiatively, such as $T^{\{cc\}}_{[\bar{u}\bar{d}]}\to D^0D^{+}\gamma$, 
$T^{\{cc\}}_{[\bar{u}\bar{s}]}\to D^0D_s^{+}\gamma$, and $T^{\{cc\}}_{[\bar{d}\bar{s}]}\to D^+D_s^{+}\gamma$.
These channels are expected to be better measured at future $Z$-factories.

In the case of $T^{\{cc\}}_{[\bar{q}\bar{q}']}$ below the $DD\gamma$ thresholds, 
they can only decay weakly and are thus stable particles. 
This is different from the exotic states discovered in experiments so far, which decay strongly. 
The scales of heavy quark weak decays, at $\mathcal{O}$(GeV), 
are different from the ones of strong decays at the scale of hundreds of MeV. 
Thus, it will provide another way to distinguish the nature of hadrons to study the weak decays of such particles. 
Since they can only weakly decay, they must be long-lived particles and cannot be kinematical effects or coupled-channel effects. 
Due to large non-perturbative contributions, it is difficult to precisely predict the branching fractions of weak decays of charmed hadrons. 
Considering the Cabibbo-favored processes with charged particles in the final states, the most favorable processes of weak decays at the LHCb are
\begin{align}
 T^{\{cc\}}_{[\bar{u}\bar{d}]}&\to D^{+}K^{-}\pi^{+},\nonumber\\
T^{\{cc\}}_{[\bar{u}\bar{s}]}&\to D_{s}^{+}K^{-}\pi^{+},\quad \text{or}\quad D^{0}\pi^{+},\quad \text{or}\quad D^{+}K^{+}
K^{-},\\
 T^{\{cc\}}_{[\bar{d}\bar{s}]}&\to D^{+}K^{+}K^{-}\pi^{+},\quad \text{or}\quad D^{+}\pi^{+}.\nonumber
 \end{align}
 
Considering a more abundant production of $T^{\{cc\}}_{[\bar{u}\bar{d}]}$ compared to the ones containg a strange quark, 
we will discuss the possible signal events of $T^{\{cc\}}_{[\bar{u}\bar{d}]}$ in experiments. 
If it decays strongly, it decays into $D^0D^{*+}$ with $D^{*+}\to D^0\pi^+$, $D^0\to K^-\pi^+$, and $K^-\pi^+\pi^+\pi^-$. 
The total branching fraction would be $\mathcal{O}(10^{-2})$. 
In case it is a stable particle, the branching fraction of $T^{\{cc\}}_{[\bar{u}\bar{d}]}\to D^{+}K^{-}\pi^{+}$ would be of the order of $10\%$. 
Considering the further decay $D^+\to K^-\pi^+\pi^+$, 
the total branching fraction of the weak decay of $T^{\{cc\}}_{[\bar{u}\bar{d}]}$ would be $\mathcal{O}(10^{-2})$ as well. 
The order of the branching fractions of $T^{\{cc\}}_{[\bar{u}\bar{d}]}$ decays is thus the same as the observed $\Xi_{cc}^{++}$. 
Comparing with the production rates between double-charm tetraquarks and baryons 
and considering approximately $2\times10^3$ events of $\Xi_{cc}^{++}$ with the current LHCb data, 
the signal yields of $T^{\{cc\}}_{[\bar{u}\bar{d}]}$ would be $\mathcal{O}(10^{2})$ at LHCb, 
either for a strongly or weakly decaying $T^{\{cc\}}_{[\bar{u}\bar{d}]}$.
This number will reach $\mathcal{O}(10^{3})$ at LHCb Run III. 
Therefore, it is expected that the double-charm tetraquark will be observed at the LHCb in the near future. 
Although the production rates are smaller at the future Z factories, 
it is also expected to be observed at the Tera-Z factories due to smaller backgrounds.

\section{Conclusion}
\label{sec:conclusion}

To search for double-charm tetraquarks $T^{\{cc\}}_{[\bar{q}\bar{q}']}$, we study their productions and decay channels at the LHCb and 
also future $Z$-factories. It is found that the LHCb has already collected plenty of events for the discovery of $T^{\{cc\}}_{[\bar{u}\bar{d}]}$. 
In case $T^{\{cc\}}_{[\bar{q}\bar{q}']}$ are above the $DD^*$ thresholds, 
we propose that the decay channel $T^{\{cc\}}_{[\bar{u}\bar{d}]}\to D^0D^{*+}\to D^0D^0\pi^+$ should be analyzed at the LHCb. 
In contrast, in case that $T^{\{cc\}}_{[\bar{q}\bar{q}']}$ are below the $DD\gamma$ thresholds, 
the decay channel $T^{\{cc\}}_{[\bar{u}\bar{d}]}\to D^{+}K^{-}\pi^{+}$ should be analyzed at the LHCb. 
Although these double-charm tetraquarks will probably be discovered at the LHCb, 
there are still some decay channels containing neutral particles in their final states,
\textit{e.g.}, $D^0D^+\gamma$, which are better to be studied at future $Z$-factories.

\section*{Acknowledgement}

The authors are grateful to Ji-Bo He and Xiao-Rui Lyu for useful discussions. 
This work is supported by Natural Science Foundation of China under grant No. 12005068 and 11975112.

\section*{Additional note}

{\small After the first version of this paper appeared on arxiv.org, with e-Print No. 2008.08026 [hep-ph], and 
during the preparation of  its submission to Chinese Physics C, the observation of the first double-charm tetraquark, 
a $T^{\{cc\}}_{[\bar{u}\bar{d}]}$-like
particle, was reported in the talks {\it "Highlights from the LHCb Experiment"} given by Franz Muheim and 
{\it "Recent LHCb results on exotic meson candidates"} given by Ivan Polyakov at the European 
Physical Society Conference on high energy physics 2021. For details, see {\it https://indico.desy.de/event/28202/contributions/102717/} 
and {\it https://indico.desy.de/event/28202/contributions/105627/}. Subsequently, the corresponding papers were released on arxiv~\cite{LHCb:2021vvq,LHCb:2021auc}.
It turns out that the observed double-charm tetraquark is an unstable particle dominated by strong decays, and indeed, 
it was discovered through the channel $T^{\{cc\}}_{[\bar{u}\bar{d}]}\to D^0D^{*+}\to D^0D^0\pi^+$, as we suggested in this paper. 
Moreover, the signal yield at the LHCb was reported to be $N=117\pm16$, which is very consistent with our prediction, $\mathcal{O}(10^2)$, using the current LHCb data.}







\end{document}